\documentclass{ws-procs9x6}
\numberwithin{equation}{section}

\begin{document}

\title{On the Notion of Proposition in Classical and Quantum Mechanics}
\author{C. Garola}

\address{Dipartimento di Fisica dell'Universit\`a and Sezione INFN, \\
Via per Arnesano, 73100 Lecce, Italy \\ 
E-mail: garola@le.infn.it}

\author{S. Sozzo}

\address{Dipartimento di Fisica dell'Universit\`a and Sezione INFN, \\ 
Via per Arnesano, 73100 Lecce, Italy \\
E-mail: sozzo@le.infn.it}  

\maketitle

\abstracts{The term \emph{proposition} usually denotes in quantum mechanics (QM) an element of (standard) quantum logic (QL). Within the orthodox interpretation of QM the propositions of QL cannot be associated with sentences of a language stating properties of individual samples of a physical system, since properties are \emph{nonobjective} in QM. This makes the interpretation of propositions problematical. The difficulty can be removed by adopting the objective interpretation of QM proposed by one of the authors (\emph{semantic realism}, or \emph{SR}, interpretation). In this case, a unified perspective can be adopted for QM and classical mechanics (CM), and a simple first order predicate calculus ${\mathcal L}(x)$ with Tarskian semantics can be constructed such that one can associate a \emph{physical proposition} (\emph{i.e.}, a set of physical states) with every sentence of ${\mathcal L}(x)$. The set ${\mathcal P}^{f}$ of all physical propositions is partially ordered and contains a subset ${\mathcal P}^{f}_{T}$ of \emph{testable physical propositions} whose order structure depends on the criteria of testability established by the physical theory. In particular, ${\mathcal P}^{f}_{T}$ turns out to be a Boolean lattice in CM, while it can be identified with QL in QM. Hence the propositions of QL can be associated with sentences of ${\mathcal L}(x)$, or also with the sentences of a suitable quantum language  ${\mathcal L}_{TQ}(x)$, and the structure of QL characterizes the notion of testability in QM. One can then show that the notion of \emph{quantum truth} does not conflict with the classical notion of truth within this perspective. Furthermore, the interpretation of QL propounded here proves to be equivalent to a previous pragmatic interpretation worked out by one of the authors, and can be embodied within a more general perspective which considers states as first order predicates of a broader language with a Kripkean semantics.}

\section{Introduction}
It is often maintained in the literature on the foundations of quantum mechanics (QM) that the lattice of propositions of quantum logic (QL)\footnote{For the sake of brevity, we simply call \emph{quantum logic} here the formal structure that is called in literature \emph{concrete}, or \emph{standard}, (\emph{sharp}) \emph{quantum logic},\cite{dcgg04} together with its standard physical interpretation.} is a logical calculus which is different from the classical logical calculus and specific of QM (see Ref. 1 for a review on this subject till the early seventies; for a more recent perspective, together with an updated bibliography, see, \emph{e.g.}, Refs. 2 and 3). Yet many scholars do not accept this view and argue that QL is a mathematical structure with a physical interpretation, not a new logic (for an explicit statement of this position see, \emph{e.g.}, Ref. 4).

In our opinion, the unsettled quarrel between the positions above finds its roots in a specific feature of the standard interpretation of QM, that is, \emph{nonobjectivity} of physical properties. Because of this feature, there are sentences attributing physical properties to samples of a given physical system that are meaningful or meaningless (\emph{i.e.}, have or have not a truth value, respectively) depending on the state of the object, and also sentences that are meaningless in any case, even if they belong to the natural language of physics (a known example of these is the statement ``the particle $x$ has position $\vec r$ and momentum $\vec p$ at time $t$''). Hence the propositions of QL cannot be connected in a direct way with sentences of this kind, following standard procedures in classical logic (CL), which makes their logical interpretation problematical (in particular, QL seems to introduce a new mysterious concept of \emph{quantum truth}\cite{bo1})\footnote{A rather recent investigation on the concept of proposition has been done by R\'{e}dei\cite{r98}. Within R\'{e}dei's analysis physical properties, or sentences about probabilities of properties, are directly taken as elementary sentences of a logical language, and propositions are identified with equivalence classes of (elementary or complex) sentences, each class containing all sentences which are equivalent with respect to a quantum concept of truth. Our analysis here considers a different kind of elementary sentences and introduces various kinds of propositions. The lattice of R\'{e}dei's propositions is then isomorphic, in QM, to the lattice of all \emph{testable physical propositions} introduced here (Sec. 6).}.

The above difficulties cannot be removed as long as nonobjectivity is maintained to be an unavoidable feature of QM. Nevertheless most physicists accept nonobjectivity, basing this acceptance on well known no--go theorems (the most famous of which are probably Bell's\cite{b64,m93} and Bell--Kochen--Specker's\cite{m93}$^{-}$\cite{ks67}). It has been proven in a number of papers by one of the authors, however, that these theorems, which are mathematically well established, rest on assumptions which follow from implicitly adopting an epistemological position which is suitable for classical physics but contrasts with the operational philosophy of QM.\cite{gs96b}$^{-}$\cite{ga05} To be precise, they assume the simultaneous validity of a set of empirical physical laws in which the observables that appear in some laws are incompatible with the observables that appear in other laws, so that it is impossible, according to QM, to check whether all the laws of the set hold simultaneously. This suggests that the simultaneous validity assumption should be dropped in QM: but, then, the no--go theorems cannot be proved. It follows that the nonobjectivity of physical properties can no more be classified as a logical necessity, but only as a (legitimate) interpretational choice, and alternative interpretations of QM in which objectivity of properties is restored become possible. An interpretation of this kind has then be constructed by one of us, together with other authors (\emph{semantic realism}, or \emph{SR}, interpretation\cite{g99}$^{-}$\cite{g02,gp04,g91,gs96a}). The SR interpretation preserves the mathematical apparatus and the statistical interpretation of QM, and yet considers every elementary sentence attributing a physical property to a given individual physical object as meaningful (though its truth value may be empirically accessible or not, depending on the state of the object).

Because of objectivity, the SR interpretation avoids the difficulties of the standard interpretation pointed out above, so that physical propositions can be introduced in QM associating them to sentences of a suitable classical predicate calculus. This allows us to propound in this paper a general scheme based on classical logic for the introduction of physical propositions in physical theories, which can then be particularized to classical mechanics (CM) and to QM. Our scheme explains, in particular, how QL can be obtained by using a testability criterion for selecting a suitable subset in the set of all physical propositions, and shows that a notion of quantum truth can be derived from the classical notion of truth as correspondence (as explicated rigorously by Tarski's semantic theory\cite{t44,t56}).

In order to favour a better understanding of the above results, let us describe the content of the present paper in more details.

In Sec. 2 we construct a classical first order predicate calculus ${\mathcal L}(x)$, with monadic predicates and one individual variable only, in which a classical (Tarskian) notion of truth is adopted, and associate a family of \emph{individual propositions}, parametrized by the interpretations of the variable, with every (open) sentence of ${\mathcal L}(x)$. 

In Sec. 3 we define \emph{physical propositions}, introduce the truth value \emph{certainly true} on ${\mathcal L}(x)$ (which adds without contradiction to the standard values \emph{true}/\emph{false}), and study some properties of the poset $({\mathcal P}^f, \subseteq)$ of all physical propositions. 

In Sec. 4 we conclude the general part of the paper by introducing the subset ${\mathcal P}_{T}^f \subseteq {\mathcal P}^f$ of all \emph{testable} physical propositions, which is basic for the analysis of measurement processes in the framework of specific physical theories (as CM and QM).

In Sec. 5 we specialize the notions introduced in the previous sections to CM. We show that, if suitable axioms (which are justified by the intended interpretation) are introduced, the concepts of individual proposition, physical proposition and testable physical proposition can be identified, which provides a very simple scheme that explains why people usually say that ``classical mechanics follows classical logic'' (which is however a misleading statement in our opinion).

In Sec. 6 we show that the different kinds of propositions introduced in the general part cannot be identified in QM, and introduce some specific axioms which are supported by the broad existing literature on QL. These allow us to construct a quantum language ${\mathcal L}_{TQ}(x)$, based on ${\mathcal L}(x)$, which is such that the set of all physical propositions associated with its sentences coincides with ${\mathcal P}_{T}^f$ and can be identified with the set of all propositions of QL. It follows that every proposition of QL can be associated with a sentence of a suitable first order predicate calculus, as in classical logic, and that the set of all propositions of QL is selected on the basis of a criterion of testability, which is tipically physical and shows the empirical character of the lattice structure of QL.

In Sec. 7 we use the interpretation provided in Sec. 6 in order to look deeper into the concept of `quantum truth'. We show that this concept directly follows in our approach from the concept of \emph{certainly true} introduced in the general part, hence it does not conflict with the classical concept of truth. This provides a satisfactory unification of notions that are usually regarded as incompatible.

In Sec. 8 we discuss the relations between the \emph{semantical} interpretation of QL provided in Sec. 6 with the \emph{pragmatic} interpretation propounded by one of us in a recent paper.\cite{g05} We show that the two interpretations can be easily translated one into the other, and that they are intuitively equivalent.

In Sec. 9 we briefly comment on our approach from a general logical perspective. We note that individual and physical propositions can be considered as propositions in a standard sense in CL if states are considered as \emph{possible worlds} (\emph{modal interpretation} of QL). This interpretation is however problematical, and we briefly sketch a possible alternative which refers to the broader language introduced by one of us, together with other authors, in some previous papers.\cite{g91,gs96a}

\section{The language ${\mathcal L}(x)$} \label{linguaggio}
The formal language that we want to construct in this section is a simplified and modified version of the more general language introduced in some previous papers\cite{g91,gs96a} with the aim of formalizing a sublanguage of the observative language of QM.

The alphabet of ${\mathcal L}(x)$ consists of an individual variable $x$, a set ${\mathcal E}= \{ E, F, \ldots \}$ of monadic predicates called \emph{properties}, a set $\{ \lnot, \land, \lor \}$ of logical connectives and a set $\{  (\, , \,) \}$ of auxiliary signs.

The formation rules for sentences, or well-formed formulas (wffs), of ${\mathcal L}(x)$ are the standard (recursive) formation rules for wffs of a classical first order predicate calculus, in which $\lnot$, $\land$, $\lor$ denote negation, conjunction and disjunction, respectively. We denote by $\phi(x)$ the set of all wffs of ${\mathcal L}(x)$, and by ${\mathcal E}(x)$ the set of all \emph{elementary} sentences (or \emph{atomic} wffs) of  ${\mathcal L}(x)$.

The semantics of ${\mathcal L}(x)$ consists of a family of Tarskian semantics parametrized by a set $\mathcal S$ of \emph{states}. Every $S \in \mathcal S$ is associated with a universe ${\mathcal U}_S$ of \emph{physical objects}. An \emph{interpretation} of the variable $x$ is a mapping $\rho:(x,S) \in \{ x \}\times {\mathcal S} \longrightarrow \rho_S(x) \in {\mathcal U}_S$. For every $S \in \mathcal S$ and $E \in \mathcal E$, an extension $ext_{S} E \subset {\mathcal U}_S$ is defined. The atomic wff $E(x)$ is \emph{true} in the state $S$ for the interpretation $\rho$ iff $\rho_S(x) \in ext_{S} E$, \emph{false} otherwise. The truth value of molecular wffs of ${\mathcal L}(x)$ is then defined following standard (recursive) truth rules in Tarskian semantics. For every interpretation $\rho$ and state $S$, we call \emph{assignment function} the mapping $\sigma_S^\rho : \phi(x) \longrightarrow \{ T,F \}$ (where $T$ stands for \emph{true} and $F$ for \emph{false}) which associates a truth value with every wff of ${\mathcal L}(x)$ following the truth rules mentioned above.

The \emph{intended interpretation} of ${\mathcal L}(x)$ is anticipated by the terminology that we have adopted. States are defined operationally as classes of physically equivalent preparation procedures (briefly, \emph{preparations}) and properties as classes of physically equivalent (ideal) registration procedures (briefly, \emph{registrations})\footnote{The notion of physical equivalence is not trivial and requires a careful analysis of the notions of preparation and (ideal) registration procedure.\cite{gs96a} We do not insist on this issue here for the sake of brevity.}. The universe ${\mathcal U}_S$ consists of samples of a prefixed physical system $\Omega$ prepared according to any preparation in $S$. Whenever an interpretation $\rho$ and a state $S$ are given, an elementary sentence, say $E(x)$, of ${\mathcal L}(x)$ states a (physical) property $E$ of the physical object $\rho_S(x) \in {\mathcal U}_S$ (by abuse of language, we often avoid mentioning the interpretation $\rho$ in the following, and briefly say that $E(x)$ attributes the property $E$ to the physical object $x$ in the state $S$).

It must be stressed that the intended interpretation of ${\mathcal L}(x)$ provided here implies that the semantics of ${\mathcal L}(x)$ is incompatible with QM whenever the standard interpretation of QM is adopted. Indeed, within this interpretation QM is maintained to be a semantically \emph{nonobjective} (or \emph{contextual}) theory, which implies that the extension $ext_S E$ is not defined for every property $E$. Hence the general scheme for propositions in physical theories propounded in this paper is based on an explicit acceptance of the SR interpretation of QM mentioned in the Introduction, which is semantically objective (we have already noted in the Introduction that the possibility of such an interpretation follows from a criticism of the implicit assumptions underlying the no--go theorems that should prove that QM is necessarily a nonobjective theory).

It must also be stressed that the operational definition of properties as classes of registrations makes every elementary wff $E(x) \in \phi(x)$ \emph{testable}, in the sense that a physical procedure exists that, under specified physical conditions, allows one to check empirically the truth value of $E(x)$. Yet, it is important to observe that this check does not reduce in all theories to registering a physical object $x$ in the state $S$ by means of a registration in $E$. There are indeed physical theories, as QM, in which the registration usually modifies the state $S$ in an unpredictable way, so that the obtained result refers to the state after the registration, not to $S$. In these theories the empirical accessibility of the truth values of $E(x)$ is then restricted to a proper subset of states which depends on $E$ (see Sec. 7).

Let us introduce now some further definitions. Firstly, two binary relations of \emph{logical preorder} $\le$ and \emph{logical equivalence} $\equiv$ can be defined on $\phi(x)$ by following standard procedures in classical logic, \emph{i.e.}, by setting, for every $\alpha(x),\beta(x)\in \phi(x)$, 
\begin{displaymath}
\alpha(x) \le \beta(x)  \quad \textrm{\emph{iff}} 
\end{displaymath}
\begin{displaymath}
\textrm{for every} \, \, \rho \in \mathcal R, \, S \in \mathcal S, 
\sigma_S^{\rho}(\alpha(x))=T \, \, \, \textrm{implies} \, \, \, \sigma_S^{\rho}(\beta(x))=T,
\end{displaymath}
and
\begin{displaymath}
\alpha(x) \equiv \beta(x) \quad \textrm{\emph{iff}} \quad \alpha(x) \le \beta(x) \, \, \textrm{and}  \, \, \beta(x) \le \alpha(x).
\end{displaymath}
It is then easy to see that the partially ordered set (briefly, \emph{poset}) \hspace*{-1mm} $(\phi(x)/_{\equiv},\hspace*{-1mm} \le)$ (where $\le$ denotes, by abuse of language, the order canonically induced on $\phi(x)/_{\equiv}$ by the preorder $\le$ defined on $\phi(x)$) is a Boolean lattice (the \emph{Lindenbaum--Tarski algebra} of ${\mathcal L}(x)$).

Secondly, let $\mathcal R$ be the set of all possible interpretations of $x$. Then, we associate an \emph{individual proposition} $p_{\alpha(x)}^{\rho}$  with every pair $(\rho, \alpha(x)) \in {\mathcal R} \times \phi(x)$, defined as follows.
\begin{equation}
p_{\alpha(x)}^{\rho}= \{ S \in {\mathcal S} \quad | \quad \sigma_S^{\rho}(\alpha(x))=T \}.
\end{equation}
The definition of $p_{\alpha(x)}^{\rho}$ implies that 
\begin{displaymath}
\sigma_S^{\rho}(\alpha(x))=T \quad \textrm{\emph{iff}} \quad S \in p_{\alpha(x)}^{\rho}.
\end{displaymath} 
Furthermore, one easily gets that, for every elementary wff $E(x) \in \phi(x)$,
\begin{equation}
p_{E(x)}^{\rho}= \{ S \in {\mathcal S} \quad | \quad \rho_S(x) \in ext_S E \},
\end{equation}
while for every $\alpha(x), \beta(x) \in \phi(x)$ one gets
\begin{eqnarray}
p_{\lnot\alpha(x)}^{\rho}= \mathcal S \setminus p_{\alpha(x)}^{\rho}, \label{not} \\
p_{\alpha(x) \land \beta(x)}^{\rho}= p_{\alpha(x)}^{\rho} \cap p_{\beta(x)}^{\rho},  \label{and} \\
p_{\alpha(x) \lor \beta(x)}^{\rho}= p_{\alpha(x)}^{\rho} \cup p_{\beta(x)}^{\rho} \label{or}
\end{eqnarray}
(where $\setminus$, $\cap$, $\cup$ denote set--theoretical subtraction, intersection and union, respectively). Let $\subseteq$ denote set--theoretical inclusion and let ${\mathcal P}^{\rho}$ be the set of all individual propositions associated with sentences of $\phi(x)$ whenever $\rho$ is fixed. Then, Eqs. (\ref{not}), (\ref{and}) and (\ref{or}) imply that also the poset $({\mathcal P}^{\rho}, \subseteq)$ is a Boolean lattice. 

Thirdly, by using the definitions of logical order, logical equivalence and proposition introduced above, we get
\begin{displaymath}
\alpha(x) \le \beta(x) \quad \textrm{\emph{iff}} \quad \textrm{for every} \, \,  \rho \in {\mathcal R}, \, p_{\alpha(x)}^{\rho} \subseteq p_{\beta(x)}^{\rho},
\end{displaymath}
\begin{displaymath}
\alpha(x) \equiv \beta(x) \quad \textrm{\emph{iff}} \quad \textrm{for every} \, \, \rho \in {\mathcal R}, \, p_{\alpha(x)}^{\rho} = p_{\beta(x)}^{\rho},
\end{displaymath}
which show that the logical relations on $\phi(x)$ imply set--theoretical relations on every set ${\mathcal P}^{\rho}$ of individual propositions.

\section{The poset of physical propositions} \label{proposizioni}
The intended interpretation of ${\mathcal L}(x)$ introduced in Sec. 2 suggests to associate a set of states with every sentence of ${\mathcal L}(x)$, to be precise the set of states which make this sentence true whatever the interpretation of the variable may be. Thus, for every $\alpha(x) \in \phi(x)$ we define a \emph{physical proposition} $p_{\alpha(x)}^{f}$, as follows.
\begin{equation} \label{pfalpha}
p_{\alpha(x)}^{f}= \{ S \in {\mathcal S} \, \, | \, \, \forall \rho \in {\mathcal R}, \, \sigma_{S}^{\rho}(\alpha(x))= T \}.
\end{equation}
By using the definitions introduced in Sec. 2, we then get
\begin{equation} \label{propositions}
p_{\alpha(x)}^{f}= \{ S \in {\mathcal S} \, \, | \, \, \forall \rho \in {\mathcal R}, \, S \in p_{\alpha(x)}^{\rho} \}= \cap_{\rho} p_{\alpha(x)}^{\rho}.
\end{equation}
We denote by ${\mathcal P}^{f}$ the set of all physical propositions associated with wffs of ${\mathcal L}(x)$, that is, we put
\begin{equation} \label{propositionset}
{\mathcal P}^{f}= \{ p_{\alpha(x)}^{f} \, \, | \, \, \alpha(x) \in \phi(x) \}.
\end{equation}
For every $\alpha(x) \in \phi(x)$ we can now introduce the notion of ``true with certainty'' by setting: 
\begin{displaymath}
\alpha(x) \, \, \textrm{is \emph{certainly true in S}} \quad \textrm{\emph{iff}} \quad S \in  p_{\alpha(x)}^{f}.
\end{displaymath}

The new notion thus follows from the standard notion of truth introduced in Sec. 2 and applies to open wffs of $\phi(x)$ independently of any interpretation of the variable $x$ (we note explicitly that we do not introduce here a notion of \emph{certainly false in S}: whenever $S \notin  p_{\alpha(x)}^{f}$, we simply say that $\alpha(x)$ is not certainly true in $S$).

The definition of physical proposition associated with a wff $\alpha(x) \in \phi(x)$ also allows us to introduce the new binary relations of \emph{physical preorder} and \emph{physical equivalence} on $\phi(x)$. For every $\alpha(x), \beta(x) \in \phi(x)$, we put
\begin{displaymath}
\alpha(x) \prec \beta(x) \quad \textrm{\emph{iff}} \quad  p_{\alpha(x)}^{f} \subseteq p_{\beta(x)}^{f},
\end{displaymath}
\begin{displaymath}
\alpha(x) \approx \beta(x) \quad \textrm{\emph{iff}} \quad  p_{\alpha(x)}^{f} = p_{\beta(x)}^{f}.
\end{displaymath}
By comparing the definitions of $\prec$ and $\approx$ with the definitions of $\le$ and $\equiv$, respectively, one gets
\begin{displaymath}
\alpha(x) \le \beta(x) \quad \textrm{\emph{implies}} \quad  \alpha(x) \prec \beta(x),
\end{displaymath}
\begin{displaymath}
\alpha(x) \equiv \beta(x) \quad \textrm{\emph{implies}} \quad  \alpha(x) \approx \beta(x),
\end{displaymath}
that is, logical preorder implies physical preorder and logical equivalence implies physical equivalence. The converse implications do not hold in general, in the sense that one cannot prove that they hold without introducing further assumptions. We come back on this issue in Secs. 5 and 6.

Let us come now to the set ${\mathcal P}^f$ of all physical propositions. This set is obviously partially ordered by set--theoretical inclusion, but the properties of the poset $({\mathcal P}^f,\subseteq)$ depend on the specific physical theory that is considered. In particular, one cannot generally assert that  $({\mathcal P}^f,\subseteq)$ is a Boolean lattice, as $({\mathcal P}^{\rho},\subseteq)$. However, some weaker features of it can be established. Indeed, let $\alpha(x),\beta(x) \in \phi(x)$. Then, the following statements hold.
\vskip 2mm

(i) $p_{\lnot\alpha(x)}^{f} \subseteq {\mathcal S} \setminus  p_{\alpha(x)}^{f}$,

(ii) $p_{\alpha(x) \land \beta(x)}^{f}=  p_{\alpha(x)}^{f} \cap  p_{\beta(x)}^{f}$,

(iii) $p_{\alpha(x) \lor \beta(x)}^{f} \supseteq  p_{\alpha(x)}^{f} \cup  p_{\beta(x)}^{f}$ 

\vskip 2mm
\noindent
(note that, generally, neither ${\mathcal S} \setminus  p_{\alpha(x)}^{f}$ nor $p_{\alpha(x)}^{f} \cup  p_{\beta(x)}^{f}$ belong to ${\mathcal P}^f$; statement (ii) shows instead that $ p_{\alpha(x)}^{f} \cap  p_{\beta(x)}^{f}$ belongs to ${\mathcal P}^f$).

Let us prove statements (i), (ii) and (iii). By using Eqs. (\ref{propositions}) and  (\ref{not}), we get
\begin{displaymath}
p_{\lnot\alpha(x)}^{f}= \cap_\rho p_{\lnot\alpha(x)}^{\rho} = \cap_{\rho} ({\mathcal S} \setminus  p_{\alpha(x)}^{\rho}) \subseteq {\mathcal S} \setminus \cap_\rho p_{\alpha(x)}^{\rho} ={\mathcal S} \setminus  p_{\alpha(x)}^{f}.
\end{displaymath}
Furthermore, by using Eqs. (\ref{propositions}) and  (\ref{and}), we get
\begin{displaymath}
p_{\alpha(x) \land \beta(x)}^{f}= \cap_\rho p_{\alpha(x) \land \beta(x)}^{\rho}= \cap_\rho (p_{\alpha(x)}^{\rho} \cap  p_{\beta(x)}^{\rho})=
\end{displaymath}
\begin{displaymath}
=( \cap_\rho p_{\alpha(x)}^{\rho}) \cap ( \cap_\rho p_{\beta(x)}^{\rho})= p_{\alpha(x)}^{f} \cap  p_{\beta(x)}^{f}.
\end{displaymath}
Finally, by using Eqs. (\ref{propositions}) and  (\ref{or}), we get
\begin{displaymath}
p_{\alpha(x) \lor \beta(x)}^{f}= \cap_\rho p_{\alpha(x) \lor \beta(x)}^{\rho}= \cap_\rho (p_{\alpha(x)}^{\rho} \cup  p_{\beta(x)}^{\rho}) \supseteq
\end{displaymath}
\begin{displaymath}
\supseteq( \cap_\rho p_{\alpha(x)}^{\rho}) \cup ( \cap_\rho p_{\beta(x)}^{\rho})= p_{\alpha(x)}^{f} \cup  p_{\beta(x)}^{f}.\end{displaymath}
To close up, we note that the definitions of $\prec$ and $\approx$ on $\phi(x)$ imply that the poset $(\phi(x)/_{\approx}, \prec)$ (where $\prec$ denotes, by abuse of language, the order canonically induced on $\phi(x)/_{\approx}$ by the preorder $\prec$ defined on $\phi(x)$) is order--isomorphic to $({\mathcal P}^f,\subseteq)$.

\section{The general notion of testability} \label{testabilita}
The intended physical interpretation of ${\mathcal L}(x)$ suggests that a sentence of ${\mathcal L}(x)$ can be classified as \emph{empirically decidable}, or \emph{testable}, iff it can be associated with a registration procedure that allows one (under physical conditions to be carefully specified, see Sec. 2) to determine its truth value whenever an interpretation $\rho$ of the variable $x$ is given. Since all elementary sentences are testable, one is thus led to define the subset $\phi_T(x) \subseteq \phi(x)$ of all \emph{testable wffs} of $\phi(x)$ as follows.
\begin{equation} \label{testable}
\phi_T(x)= \{ \alpha(x) \in \phi(x) \, \, | \, \, \exists E_{\alpha} \in {\mathcal E} \, : \, \alpha(x) \equiv E_{\alpha}(x) \}. 
\end{equation} 
The subset ${\mathcal P}_{T}^{f} \subseteq {\mathcal P}^{f}$ of all physical propositions associated with wffs of $\phi_T(x)$ will then be called \emph{the set of all testable physical propositions}. More formally,
\begin{equation}
{\mathcal P}_{T}^{f}= \{ p_{\alpha(x)}^{f}\in {\mathcal P}^{f} \, \, | \, \, \alpha(x) \in \phi_T(x) \}.
\end{equation} 
Of course, $\phi_T(x)$ is preordered by the restrictions of the preorders $\le$ and $\prec$ defined on $\phi(x)$ to it. For the sake of simplicity, we will denote preorders and equivalence relations on $\phi_T(x)$ by the same symbols used to denote them on $\phi(x)$. Hence, the logical preorder $\le$ implies the physical preorder $\prec$, and the logical equivalence  $\equiv$ implies the physical equivalence $\approx$ also on $\phi_T(x)$. We thus get two preorder structures, $(\phi_T(x), \le)$ and $(\phi_T(x), \prec)$, and two posets $(\phi_T(x)/_{\equiv}, \le)$ and $(\phi_T(x)/_{\approx}, \prec)$. The latter, in particular,  is isomorphic to $({\mathcal P}_{T}^{f}, \subseteq)$.

We shall see in the next sections some further characterizations of the foregoing posets within the framework of specific theories.

\section{Classical mechanics (CM)} \label{meccanica classica}
It is well known that in classical mechanics (CM) all physical objects in a given state $S$ possess the same properties. This feature of CM can be formalized here by introducing the following assumption.

\bigskip
\noindent
\textbf{CMS.} \emph{For every $S \in \mathcal S$ and $E \in \mathcal E$, either $ext_{S} E={\mathcal U}_S$ or  $ext_{S} E=\emptyset$.} 

\bigskip
\noindent
It follows from assumption CMS that, for every interpretation $\rho \in \mathcal R$, $\rho_{S}(x)\in ext_S E$ iff $ext_{S} E={\mathcal U}_S$, and $\rho_{S}(x)\notin ext_S E$ iff $ext_{S} E=\emptyset$. Therefore, the assignment function $\sigma_S^{\rho}$ does not depend on the specific interpretation $\rho$. More explicitly, for every interpretation $\rho$ and state $S$,
\begin{displaymath}
\left \{ \begin{array}{c}
\begin{tabular}{ccc}
$\sigma^{\rho}_S(E(x))=T$ & \emph{iff} & $ext_S E= {\mathcal U}_S$ \\
$\sigma^{\rho}_S(E(x))=F$ & \emph{iff} & $ext_S E=\emptyset$
\end{tabular}
\end{array} \right.,
\end{displaymath}
\begin{displaymath}
\left \{ \begin{array}{c}
\begin{tabular}{ccc}
$\sigma^{\rho}_S(E(x) \land F(x))=T$ & \emph{iff} & $ext_S E= {\mathcal U}_S=ext_S F$ \\
$\sigma^{\rho}_S(E(x) \land F(x))=F$ & \emph{iff} & $ext_S E \ne  ext_S F$
\end{tabular}
\end{array} \right.
\end{displaymath}
(where $E,F \in \mathcal E$), etc.

Since $\sigma_S^{\rho}$ does not depend on $\rho$, neither the individual proposition $p_{\alpha(x)}^{\rho}$ depends on $\rho$, and we can omit writing the index $\rho$ in both symbols. Thus, for every $\rho \in \mathcal R$, the individual proposition associated with $\alpha(x) \in \phi(x)$ is given by
\begin{equation}
p_{\alpha(x)}= \{ S \in {\mathcal S} : \sigma_{S}(\alpha(x))=T \}. 
\end{equation} 
More explicitly, we have
\begin{equation}
p_{E(x)}= \{ S \in {\mathcal S} : ext_S E= {\mathcal U}_{S} \},
\end{equation}
\begin{equation}
p_{E(x) \land F(x)}= \{ S \in {\mathcal S} : ext_S E= {\mathcal U}_{S}= ext_S F \}= p_{E(x)} \cap p_{F(x)}.
\end{equation}
etc.

The set ${\mathcal P}^{\rho}$ of all individual propositions associated with wffs of ${\mathcal L}(x)$ obviously does not depend on $\rho$, and will be simply denoted by $\mathcal P$. Because of the above specific features, the general notions introduced in Secs. 2, 3, 4 particularize in CM as follows. 

For every $\alpha(x), \beta(x) \in \phi(x)$, and $S \in \mathcal S$,
\begin{center}
\begin{tabular}{lll}
$\sigma_S(\alpha(x))=T$ & \emph{iff} & $\quad S \in p_{\alpha(x)}$, \\
$\alpha(x) \le \beta(x)$ & \emph{iff} & $p_{\alpha(x)} \subseteq p_{\beta(x)}$, \\ 
$\alpha(x) \equiv \beta(x)$ & \emph{iff} &  $p_{\alpha(x)} = p_{\beta(x)}$. 
\end{tabular}
\end{center}
It also follows from the general case that the Lindenbaum--Tarski algebra $(\phi(x)/_{\equiv}, \le)$ of ${\mathcal L}(x)$ is isomorphic to the Boolean lattice of individual propositions $({\mathcal P}, \subseteq)$, so that the two lattices can be identified.

Coming to physical propositions, we get, for every $\alpha(x) \in \phi(x)$,
\begin{equation}
p_{\alpha(x)}^f=p_{\alpha(x)},
\end{equation}
and, therefore, ${\mathcal P}^f=\mathcal P$. Thus, the set of all physical propositions coincides in CM with the set of all individual propositions, and the notions of \emph{true} and \emph{certainly true} also coincide.

Furthermore the intended physical interpretation suggests that every sentence of the language ${\mathcal L}(x)$ is testable in CM. This inspires the following assumption.

\bigskip
\noindent
\textbf{CMT.} \emph{The set of all testable sentences of the language ${\mathcal L}(x)$ coincides in CM with the set of all sentences of ${\mathcal L}(x)$, that is, $\phi_T(x)=\phi(x)$ in CM.}

\bigskip
\noindent
Assumption CMT implies that ${\mathcal P}_T^f={\mathcal P}^f=\mathcal P$, whence
\begin{displaymath}
({\mathcal P}_{T}^f, \subseteq)= ({\mathcal P}, \subseteq).
\end{displaymath}
More explicitly, the poset of all testable physical propositions of a physical system $\Omega$ coincides with the poset of all individual propositions of its language ${\mathcal L}(x)$, and has the structure of a Boolean lattice. This result explains, in particular, the common statement in the literature that ``the logic of a classical mechanical system is a classical propositional logic''.\cite{r98} This statement is however misleading in our opinion, since it ignores the conceptual difference between individual, physical and testable physical propositions, that coincide in CM only because of assumptions CMS and CMT.

\section{Quantum mechanics (QM)} \label{meccanica quantistica}
We have stressed in Sec. 2 that our semantics (hence the general scheme in Secs. 2, 3 and 4) is unsuitable for QM whenever the standard interpretation of this theory is accepted. As anticipated in the Introduction and in Sec. 2, we therefore adopt in the present paper the SR interpretation of QM worked out by one of the authors and by other authors in a series of articles,\cite{g99}$^{-}$\cite{g02,gp04,g91,gs96a} according to which $ext_{S} E$ can be defined in every physical situation (we show in Sec. 7 that the new perspective also allows us to elucidate the concept of quantum truth underlying the standard interpretation of QM). At variance with CM, it may then occur in QM that $\emptyset \ne ext_{S} E \ne {\mathcal U}_S$, so that the assignment function $\sigma_{S}^{\rho}$ generally depends on the interpretation $\rho$. The formulas written down for the general case cannot be simplified as in Sec. 5. In particular, ${\mathcal P}^{f} \ne {\mathcal P}^{\rho}$, assumptions CMS and CMT do not hold, and ${\mathcal P}_{T}^{f} \subset {\mathcal P}^{f}$.

In order to discuss how the general case particularizes when QM is considered, let us briefly remind the mathematical representations of physical systems, states and properties within this theory.

Let $\Omega$ be a physical system. Then, $\Omega $ is associated with a separable Hilbert space $\mathcal H$ over the field of complex numbers. Let us denote by $(\mathcal{L(H)},\subseteq)$ the poset of all closed subspaces of $\mathcal H $, partially ordered by set--theoretical inclusion, and let ${\mathcal A} \subset \mathcal{L(H)}$ be the set of all one--dimensional subspaces of $\mathcal H $. Then (in absence of superselection rules) a mapping
\begin{equation} \label{statiatomi}
\varphi:S\in {\mathcal S} \longrightarrow \varphi (S)\in \mathcal{A}
\end{equation}
exists which maps bijectively the set $\mathcal{S}$ of all pure states of $\Omega$ onto $\mathcal{A}$ (for the sake of simplicity, we will not consider mixed states in this paper, so that we understand the word \emph{pure} in the following)\footnote{It follows easily that every pure state S can also be represented by any vector $|\psi \rangle \in \varphi (S)\in \mathcal{A}$, which is the standard representation adopted in elementary QM. Moreover, a pure state $S$ is usually represented by an (orthogonal) projection operator on $\varphi (S)$ in more advanced QM. However, the representation $\varphi $ introduced here is more suitable for our purposes in the present paper.}. In addition, a mapping
\begin{equation} \label{proprietaproiettori}
\chi :E\in \mathcal{E\longrightarrow \chi }(E)\in \mathcal{L(H)}
\end{equation}
exists which maps bijectively the set $\mathcal E$ of all properties of $\Omega $ onto $\mathcal{L(H)}$.

The poset $(\mathcal{L(H)},\subseteq)$ is characterized by a set of mathematical properties. In particular, it is a complete, orthocomplemented, weakly modular, atomic lattice which satisfies the covering law.\cite{m63}$^{-}$\cite{bc81} We denote by $^{\bot }$, $\Cap $ and $\Cup $ orthocomplementation, meet and join, respectively, in $(\mathcal{L(H)}, \subseteq)$ (it is important to observe that $\Cap $ coincides with the set--theoretical intersection $\cap $ of subspaces of $\mathcal{L(H)}$, while  $^{\bot }$ does not generally coincide with the set--theoretical complementation $^{\prime}$, nor $\Cup$ coincides with the set--theoretical union $\cup $). Furthermore, we note that $\mathcal{A}$ obviously coincides with the set of all atoms of $(\mathcal{L(H)},\subseteq)$.

Let us denote by $\prec $ the order induced on $\mathcal{E}$, via the bijective representation $\chi $, by the order $\subseteq$ defined on $\mathcal{L(H)}$. Then, the poset $(\mathcal{E},\prec )$ is order--isomorphic to $(\mathcal{L(H)},\subseteq)$, hence it is characterized by the same mathematical properties characterizing $(\mathcal{L(H)},\subseteq)$. In particular, the unary operation induced on it, via $\chi $, by the orthocomplementation defined on $(\mathcal{L(H)},\subseteq)$, is an orthocomplementation, and $(\mathcal{E},\prec )$ is an orthomodular (\emph{i.e.}, orthocomplemented and weakly modular) lattice, usually called \emph{the lattice of properties} of $\Omega $. By abuse of language, we denote the lattice operations on $(\mathcal{E},\prec )$ by the same symbols used above in order to denote the corresponding lattice operations on $(\mathcal{L(H)},\subseteq)$.

Orthomodular lattices are said to characterize semantically \emph{orthomodular QLs }in the literature.\cite{dcgg04} The lattice of properties $(\mathcal{E},\prec )$ is a less general structure in QM, since it inherits a number of further properties from $(\mathcal{L(H)},\subseteq)$, and can be identified with the concrete, or standard, sharp QL mentioned in Sec. 1 (simply called QL here for the sake of brevity).

A further lattice, isomorphic to $(\mathcal{E},\prec)$, will be used in the following. In order to introduce it, let us consider the mapping
\begin{equation}
\theta :E\in \mathcal{E}\longrightarrow \mathcal{S}_{E}=\{S\in \mathcal{S}
\mid \varphi (S)\subseteq \chi (E)\}\in \mathcal{L(S)},
\end{equation}
where $\mathcal{L(S)}=\{\mathcal{S}_{E}\mid E\in \mathcal{E}\}$ is the range of $\theta$, and generally is a proper subset of the power set $\mathcal{P(S)}$ of $\mathcal{S}$. The poset $(\mathcal{L(S)},\subseteq )$ is order--isomorphic to $(\mathcal{L(H)},\subseteq )$, hence to $(\mathcal{E},\prec )$, since $\varphi $ and $\chi $ are bijective, so that $\theta$ is bijective and order--preserving. Therefore $(\mathcal{L(S)},\subseteq)$ is characterized by the same mathematical properties characterizing $(\mathcal{E},\prec )$. In particular, the unary operation induced on it, via $\theta$, by the orthocomplementation defined on $(\mathcal{E},\prec )$, is an orthocomplementation, and $(\mathcal{L(S)},\subseteq )$ is an orthomodular
lattice. We denote orthocomplementation, meet and join on $(\mathcal{L(S)},\subseteq )$ by the same symbols $^{\bot }$, $\Cap $, and $\Cup $, respectively, that we have used in order to denote the corresponding operations on $(\mathcal{L(H)},\subseteq )$ and $(\mathcal{E},\prec )$, and call $(\mathcal{L(S)},\subseteq )$ \emph{the lattice of closed subsets of} $\mathcal{S}$ (the word \emph{closed} refers here to the fact that, for every $\mathcal{S}_{E}\in $ $\mathcal{L(S)}$, $(\mathcal{S}_{E}^{\bot})^{\bot }=\mathcal{S}_{E}$). We also note that the operation $\Cap $ coincides with the set--theoretical intersection $\cap $ on $\mathcal{L(S)}$ because of the analogous result holding in $(\mathcal{L(H)},\subseteq)$\footnote{Whenever the dimension of $\mathcal H$ is finite, the lattice $(\mathcal{L(H)},\subseteq)$ and/or the lattice  $(\mathcal{L(S)},\subseteq)$ can be identified with Birkhoff and von Neumann's modular lattice of \emph{experimental propositions}, which was introduced in the 1936 paper that started the research on QL.\cite{bvn36} This identification is impossible if the dimension of $\mathcal H$ is not finite, since $(\mathcal{L(H)},\subseteq)$ and $(\mathcal{L(S)},\subseteq)$ are weakly modular but not modular in this case. Birkhoff and von Neumann's requirement of modularity has deep roots in von Neumann's concept of probability in QM according to some authors.\cite{r98}}.
  

Basing on the above definitions, we now introduce the following assumption. 

\bigskip
\noindent
\textbf{QMT.} \emph{The poset $({\mathcal P}_{T}^{f}, \subseteq)$ of all testable physical propositions associated with statements of $\phi_{T}(x)$ (equivalently, with atomic statements of ${\mathcal L}(x)$) coincides in QM with the lattice $({\mathcal L}({\mathcal S}), \subseteq)$ of all closed subsets of  $\mathcal S$.}

\bigskip
\noindent
Assumption QMT is intuitively natural, and can be justified by using the standard statistical interpretation of QM. We do not insist on this topic here for the sake of brevity. We note instead that assumption QMT implies that the posets $(\phi_{T}(x)/_{\approx} , \prec)$ and $({\mathcal P}_{T}^{f}, \subseteq)$, on one side, and the lattices $({\mathcal L}({\mathcal S}), \subseteq)$,  $(\mathcal{L(H)},\subseteq)$, $({\mathcal E}, \prec)$ on the other side, are order--isomorphic. Therefore also the operations of meet, join and orthocomplementation on $(\phi_{T}(x)/_{\approx} , \prec)$ and $({\mathcal P}_{T}^{f}, \subseteq)$ will be denoted by the symbols $\doublecap$, $\doublecup$ and $^{\perp}$, respectively. The link of these operations with set--theoretical meet, join and complementation in the set $\mathcal S$ of all states can be established as follows. For every $\alpha(x),\beta(x) \in \phi_{T}(x)$,
\begin{eqnarray}
(p_{\alpha(x)}^{f})^{\perp} \subseteq {\mathcal S} \setminus p_{\alpha(x)}^{f}, \label{lnot} \\
p_{\alpha(x)}^{f} \cap p_{\beta(x)}^{f}=p_{\alpha(x)}^{f} \doublecap p_{\beta(x)}^{f}, \label{land} \\
p_{\alpha(x)}^{f} \cup p_{\beta(x)}^{f} \subseteq p_{\alpha(x)}^{f} \doublecup p_{\beta(x)}^{f} \label{lor}.
\end{eqnarray}
The isomorphisms above allow one to recover QL as a quotient algebra of sentences of ${\mathcal L}(x)$. They, however, make intuitively clear that associating the properties (or `propositions') of QL with sentences of ${\mathcal L}(x)$ is not trivial. The association requires indeed selecting testable wffs of $\phi(x)$, grouping them into classes of physical rather than logical equivalence, adopting assumption QMT, and, finally, identifying $({\mathcal L}({\mathcal S}), \subseteq)$ with $({\mathcal E}, \prec)$.

The isomorphisms above also suggest looking deeper into the links existing between the logical operations defined on $\phi(x)$ and the lattice operations of QL. To this end, let us note that statements (i), (ii) and (iii) in Sec. 3, if compared with Eqs. (\ref{lnot}), (\ref{land}) and (\ref{lor}), respectively, yield, for every $\alpha(x), \beta(x) \in \phi_T(x)$,
\begin{eqnarray}
p_{\lnot\alpha(x)}^{f} \subseteq {\mathcal S} \setminus  p_{\alpha(x)}^{f} \supseteq ( p_{\alpha(x)}^{f})^{\perp}, \label{qnot} \\
p_{\alpha(x) \land \beta(x)}^{f}=  p_{\alpha(x)}^{f} \cap  p_{\beta(x)}^{f}= p_{\alpha(x)}^{f} \Cap  p_{\beta(x)}^{f}, \label{qand} \\
p_{\alpha(x) \lor \beta(x)}^{f} \supseteq  p_{\alpha(x)}^{f} \cup  p_{\beta(x)}^{f} \subseteq  p_{\alpha(x)}^{f} \Cup  p_{\beta(x)}^{f}. \label{qor} 
\end{eqnarray}
Eq. (\ref{qand}) shows that, if $\alpha(x)$ and $\beta(x)$ belong to $\phi_T(x)$, then $\alpha(x) \land \beta(x)$ belongs to $\phi_T(x)$, and establishes a strong connection between the connective $\land$ of ${\mathcal L}(x)$ and the lattice operation $\Cap$ of QL. Eqs. (\ref{qnot}) and (\ref{qor}) establish instead only weak connections between the connectives $\lnot$ and $\lor$, from one side, and the lattice operations $^{\perp}$ and $\Cup$, from the other side. Hence, no simple structural correspondence can be established between ${\mathcal L}(x)$ and QL.

One can, however, obtain a more satisfactory correspondence between the sentences of a suitable language and the `propositions' of QL by using a fragment of ${\mathcal L}(x)$ in order to construct a new quantum language ${\mathcal L}_{TQ}(x)$, as follows.

First of all, we consider two properties $E,F \in \mathcal E$ and observe that, since the mapping $\chi$ introduced in Eq. (\ref{proprietaproiettori}) is bijective, $E$ and $F$ coincide whenever they are represented by the same subspace of ${\mathcal L}({\mathcal H})$. This implies that the following sequence of equivalences holds.
\begin{displaymath}
p_{E(x)}^{f}=p_{F(x)}^{f} \quad \textrm{\emph{iff}} \quad E=F \quad \textrm{\emph{iff}} \quad E(x) \approx F(x) \quad \textrm{\emph{iff}} \quad E(x) \equiv F(x).
\end{displaymath}
It follows in particular that every equivalence class of $\phi_{T}(x)/_{\approx}$ contains one and only one atomic wff of ${\mathcal L}(x)$. Since the set  ${\mathcal E}(x)$ of all atomic wffs of  ${\mathcal L}(x)$ (Sec. 2) belongs to $\phi_T(x)$, we conclude that the correspondence that maps every $\alpha(x) \in \phi_T(x)$ onto the atomic wff $E_{\alpha}(x)$, the existence of which is guaranteed by Eq. (\ref{testable}), is a surjective mapping. Moreover, this mapping maps all physically equivalent wffs of $\phi_T(x)$ onto the same atomic wff of ${\mathcal E}(x)$.

Secondly, let us consider the set $\phi_{\land}(x)$ of all wffs of ${\mathcal L}(x)$ which either are atomic or contain the connective $\land$ only. Because of Eq. (\ref{qand}), the proposition associated with a wff $\alpha(x) \land \beta(x)$ of this kind belongs to ${\mathcal P}_T^{f}$, hence $\alpha(x) \land \beta(x)$ belongs to $\phi_T(x)$, so that $\phi_{\land}(x) \subseteq \phi_T(x)$. Then, let us introduce a new connective $\lnot_{Q}$ (\emph{quantum negation}) which can be applied (repeatedly) to wffs of $\phi_{\land}(x)$ following standard formation rules for negation connectives. We thus obtain a new formal language ${\mathcal L}_{TQ}(x)$, whose set of wffs will be denoted by $\phi_{TQ}(x)$. We adopt the semantic rules introduced in Sec. 2 for all wffs of $\phi_{\land}(x)\subseteq \phi_{TQ}(x)$, and complete the semantics of ${\mathcal L}_{TQ}(x)$ by means of the following rule.

\bigskip
\noindent
\textbf{QN.} \emph{Let $\alpha(x)\in \phi_{TQ}(x)$ and let a wff $E_{\alpha}(x) \in {\mathcal E}(x)$ exist such that $\alpha(x)$ is true} iff \emph{$E_{\alpha}(x)$ is true. Then, $\lnot_{Q} \alpha(x)$ is true} iff \emph{$E_{\alpha}^{\perp}(x)$ is true.}

\bigskip
\noindent
It is easy to see that rule QN implies that for every $\alpha(x) \in \phi_{TQ}(x)$, an elementary wff $E_{\alpha}(x)$ exists such that $\alpha(x)$ is true iff $E_{\alpha}(x)$ is true. This conclusion has the following immediate consequences. 

(i) One can define, for every interpretation $\rho$ of the variable $x$ and state $S$, an assignment function $\tau_{S}^{\rho}:\phi_{TQ}(x) \longrightarrow \{ T,F \}$. Hence, a logical preorder and a logical equivalence relation (that we still denote by the symbols $\le$ and $\equiv$, respectively, by abuse of language) can be defined on $\phi_{TQ}(x)$ by using the definitions in Sec. 2 with $\phi_{TQ}(x)$ in place of $\phi(x)$ and $\tau_{S}^{\rho}$ in place of $\sigma_{S}^{\rho}$.

(ii) One can associate a physical proposition with every $\alpha(x) \in \phi_{TQ}(x)$ by using Eq. (\ref{pfalpha}) with $\tau_{S}^{\rho}$ in place of $\sigma_{S}^{\rho}$. Hence a physical preorder and a physical equivalence relation (that we still denote by the symbols $\prec$ and $\approx$, respectively, by abuse of language) can be defined on $\phi_{TQ}(x)$ by using the definitions in Sec. 3 with $\phi_{TQ}(x)$ in place of $\phi(x)$ (one can also show that $\approx$ coincides with $\equiv$ on $\phi_{TQ}(x)$).

(iii) The notion of testability introduced in Sec. 4 can be extended to ${\mathcal L}_{TQ}(x)$ by using Eq. (\ref{testable}) with $\phi_{TQ}(x)$ in place of $\phi(x)$, obtaining that all wffs of $\phi_{TQ}(x)$ are testable. Hence, the set of all physical propositions associated with wffs of $\phi_{TQ}(x)$ coincides with ${\mathcal P}_T^{f}$.

It follows from (ii) and (iii) that $(\phi_{TQ}(x)/_{\approx}, \prec)$ is isomorphic to the lattice $({\mathcal P}_T^{f}, \subseteq)$, so that these two order structures can be identified.

The set of connectives defined on ${\mathcal L}_{TQ}(x)$ can now be enriched by introducing derived connectives. In particular, a \emph{quantum join} can be defined by setting, for every $\alpha(x),\beta(x) \in \phi_{TQ}(x)$, 
\begin{equation}
\alpha(x) \lor_Q \beta(x)=\lnot_{Q}(\lnot_{Q}\alpha(x) \land \lnot_{Q}\beta(x)).
\end{equation}
It is then easy to show that the following equalities hold.
\begin{eqnarray}
p_{\lnot_{Q} \alpha(x)}^{f}=(p_{\alpha(x)}^{f})^{\perp}, \\
p_{\alpha(x) \land \beta(x)}^{f}=p_{\alpha(x)}^{f} \doublecap p_{\beta(x)}^{f}, \\  
p_{\alpha(x) \lor_{Q} \beta(x)}^{f}=p_{\alpha(x)}^{f} \doublecup p_{\beta(x)}^{f}.
\end{eqnarray}
The equations above establish a strong connection between the logical operations defined on $\phi_{TQ}(x)$ and the lattice operations of QL. Hence, a structural correspondence exists between ${\mathcal L}_{TQ}(x)$ and QL, and the latter can be recovered within our general scheme also by firstly considering the set of all elementary wffs of ${\mathcal L}(x)$, and then constructing ${\mathcal L}_{TQ}(x)$ and the quotient algebra $(\phi_{TQ}(x)/_{\approx}, \prec)$. It is now apparent that \emph{the semantic rules for quantum connectives have an empirical character} (they depend on the mathematical representation of states and properties in QM and on assumption QMT) and that \emph{they coexist with the semantic rules for classical connectives in our approach} (the deep reason of this is, of course, our adoption of the SR interpretation of QM). In our opinion, these conclusions are relevant, since they deepen and formalize a new perspective on QL that has been propounded in some previous papers\cite{gs96b}$^{-}$\cite{gs96a} and is completely different from the standard viewpoint about this kind of logic.

To conclude, let us observe that a further derived connective $\xrightarrow[Q]{}$ can be introduced in $\phi_{TQ}(x)$ by setting, for every $\alpha(x),\beta(x)\in \phi_{TQ}(x)$,
\begin{equation}
\alpha(x)\xrightarrow[Q]{} \beta(x)=(\lnot_{Q} \alpha(x)) \lor_{Q} (\alpha(x) \land \beta(x)).  
\end{equation}  
One can thus recover within ${\mathcal L}_{TQ}(x)$ the \emph{Sasaki hook}, the role of which is largely discussed in the literature on QL.\cite{r98,dcgg04,bc81}

\section{Quantum truth}
The general notion of \emph{certainly true} introduced in Sec. 3 is defined for all wffs of ${\mathcal L}(x)$. Yet, according to our approach, only wffs of $\phi_{T}(x)$ can be associated with empirical procedures which allow one to check whether they are certainly true or not. Whenever $\alpha(x) \in \phi_{T}(x)$, the notion of \emph{certainly true} can be worked out in order to define a verificationist notion of \emph{quantum truth} (\emph{Q--truth}) in QM, as follows.

\bigskip
\noindent
\textbf{QT}. \emph{Let $\alpha(x) \in \phi_{T}(x)$. Then, we put:}

\emph{$\alpha(x)$ is} Q--true \emph{in $S \in \mathcal S$} iff  \emph{$S \in p_{\alpha(x)}^{f}$;} 

\emph{$\alpha(x)$ is} Q--false \emph{in $S \in \mathcal S$} iff \emph{$S \in (p_{\alpha(x)}^{f})^{\perp}$;} 

\emph{$\alpha(x)$ has no Q--truth value in  $S \in \mathcal S$ (equivalently, $\alpha(x)$ is} Q--indeterminate \emph{in $S$)} iff  \emph{$S \in {\mathcal S} \setminus (p_{\alpha(x)}^{f} \cup (p_{\alpha(x)}^{f})^{\perp})$}.

\bigskip
\noindent
It obviously follows from definition QT that $\alpha(x)$ is Q--true in $S$ iff it is certainly true in $S$.

Definition QT can be physically justified by using the analysis of the notion of truth in QM recently provided by ourselves\cite{gs04} and successively deepened by one of us.\cite{g05} We only note here that it is equivalent to defining a wff $\alpha(x)\in \phi(x)$ as Q--true (Q--false) in $S$ iff:

(i) $\alpha(x)$ is testable;

(ii) $\alpha(x)$ can be tested and found to be true (false) on the physical object $x$ without altering the state $S$ of $x$. 

The proof of the equivalence of the two definitions is rather simple but requires some use of the laws of QM (see again Refs. 21 and 26).

It is apparent that the notions of truth and Q--truth coexist in our approach. Indeed, a wff $\alpha(x) \in \phi(x)$ is Q--true (Q--false) for a given state $S$ of the physical system iff it belongs to $\phi_{T}(x)$ and it is true (false) independently of the interpretation of the variable $x$ (equivalently, iff it belongs to $\phi_{T}(x)$ and can be empirically proved to be true or false without altering the state $S$ of $x$). This realizes an \emph{integrated perspective}, according to which the classical and the quantum conception of truth are not mutually incompatible.\cite{g05,gs04,g06} However, definition QT introduces the notion of Q--truth on a fragment only (the set $\phi_T(x) \subset \phi(x)$) of the language ${\mathcal L}(x)$. If one wants to introduce this notion on the set of all wffs of a suitable quantum language, one can refer to the language ${\mathcal L}_{TQ}(x)$ constructed at the end of Sec. 6. Then, all wffs of $\phi_{TQ}(x)$ are testable, and definition QT can be applied in order to define Q--truth on ${\mathcal L}_{TQ}(x)$ by simply substituting  $\phi_{TQ}(x)$ to  $\phi_{T}(x)$ in it. Again, classical truth and Q-truth may coexist on ${\mathcal L}_{TQ}(x)$ in our approach.

Let us close this section by commenting briefly on the notion of truth within standard interpretation of QM. Whenever this interpretation is adopted, the languages ${\mathcal L}(x)$ and ${\mathcal L}_{TQ}(x)$ can still be formally introduced, but no classical semantics can be defined on them because of the impossibility of defining, for every $S \in \mathcal S$ and $E \in \mathcal E$, $ext_{S} E$ (see Sec. 2). One can still define, however, a notion of Q--truth for ${\mathcal L}_{TQ}(x)$. Indeed, one can firstly introduce a mapping $\chi: \alpha(x) \in \phi_{TQ}(x) \longrightarrow E_{\alpha} \in \mathcal E$ by means of recursive rules, as follows.
\begin{displaymath}
\begin{tabular}{ll}
For every $\alpha(x) \in \phi_{TQ}(x)$, & $\chi(\lnot_{Q}\alpha(x))=E_{\alpha}^{\perp}$, \\
For every $\alpha(x),\beta(x) \in \phi_{TQ}(x)$, & $\chi(\alpha(x) \land \beta(x)=E_{\alpha} \Cap E_{\beta}$. \\
\end{tabular}
\end{displaymath}
Then, one can associate a physical proposition $p_{\alpha(x)}^{f} \in {\mathcal L}({\mathcal S})$ with every $\alpha(x) \in \phi_{TQ}(x)$ by setting  $p_{\alpha(x)}^{f}=\theta(E_{\alpha})$. Finally, one can define Q--truth on $\phi_{TQ}(x)$ by means of definition QT, independently of any classical definition of truth. 

It is apparent that the above notion of Q--truth can be identified with the (verificationist\cite{gs04}) quantum notion of truth whose peculiar features have been widely explored by the literature on QL (in particular, a \emph{tertium non datur} principle does not hold in ${\mathcal L}_{TQ}(x)$). Hence, the interpretation of QL as a new way of reasoning which is typical of QM seems legitimate. But this widespread opinion is highly problematical. Indeed, whenever $S$ is given, some wffs of $\phi_{TQ}(x)$ have a truth value, some have not, quantum connectives are not truth--functional and the notion of truth appears rather elusive and mysterious.\cite{bo1} Accepting our general perspective provides instead a reinterpretation of the notion of truth underlying the standard interpretation of QM, reconciling it with classical truth, and allows one to avoid the paradoxes following from the simultaneous (usually implicit) adoption of two incompatible notions of truth (classical and quantum). 

\section{The pragmatic interpretation of QL}
The definition of \emph{Q--true in $S$} as \emph{certainly true in $S$} for wffs of $\phi_{T}(x)$ in Sec. 7 suggests, intuitively, that the assertion of a sentence $\alpha(x)$ of $\phi_{T}(x)$ should be considered justified in $S$ whenever $\alpha(x)$ is Q--true in $S$, unjustified otherwise. This informal definition can be formalized by introducing the assertion sign $\vdash$ and setting
\begin{center}
$\vdash \alpha(x)$ is \emph{justified} (\emph{unjustified}) in $S$ \emph{iff} \\
$\alpha(x)$ is Q--true (not Q--true) in $S$. 
\end{center}
The set of all elementary wffs of $\phi_{T}(x)$, each preceded by the assertion sign $\vdash$, can be identified with the set of all elementary assertive formulas of the quantum pragmatic language ${\mathcal L}_{Q}^{P}$ introduced by one of the authors in a recent paper\cite{g05} in order to provide a pragmatic interpretation of QL\footnote{It must be noted that the pragmatic interpretation of QL has some advantages with respect to the interpretation propounded in Sec. 6. In particular, it is independent of the interpretation of QM that is accepted (standard or SR), while our interpretation in this paper follows from adopting a classical notion of truth, hence from accepting the SR interpretation of QM.}. The set $\psi_{A}^{Q}$ of all \emph{assertive formulas} (afs) of ${\mathcal L}_{Q}^{P}$ is made up by all aforesaid elementary afs plus all formulas obtained by applying recursively the \emph{pragmatic connectives} $N$, $K$, $A$ to elementary afs. For every $S \in \mathcal S$ a \emph{pragmatic evaluation function} $\pi_S$ is defined which assigns a \emph{justification value} (justified/unjustified) to every af of $\psi_{A}^{Q}$ and allows one to introduce on $\psi_A^Q$ a preorder $\prec$ and an equivalence relation $\approx$ following standard procedures. More important, a \emph{p--decidable sublanguage} ${\mathcal L}_{QD}^{P}$ of ${\mathcal L}_{Q}^{P}$ can be constructed whose set $\phi_{AD}^{Q}$ of afs consists of a suitable subset of all afs of $\psi_{A}^{Q}$ which have a justification value that can be determined by means of empirical procedures of proof (in particular, all elementary afs of $\psi_A^{Q}$ belong to $\phi_{AD}^{Q}$). ${\mathcal L}_{QD}^{P}$ can then be compared with the quantum language ${\mathcal L}_{TQ}(x)$ introduced at the end of Sec. 6 by constructing a one--to--one mapping $\tau$ of $\phi_{TQ}(x)$ onto $\phi_{AD}^{Q}$, as follows.
\begin{displaymath}
\begin{tabular}{ll}
For every $E(x) \in \phi_{TQ}(x)$, & $\tau (E(x))=\vdash E(x)$, \\
For every $\alpha (x) \in \phi_{TQ}(x)$, & $\tau (\lnot_{Q} \alpha(x))=N\vdash \alpha(x)$, \\
For every $\alpha (x), \beta(x) \in \phi_{TQ}(x)$, & $\tau(\alpha(x) \land \beta(x))=\vdash \alpha(x) K \vdash \beta(x)$, \\
For every $\alpha (x), \beta(x) \in \phi_{TQ}(x)$, & $\tau(\alpha(x) \lor_{Q} \beta(x))=\vdash \alpha(x) A \vdash \beta(x)$.
\end{tabular}
\end{displaymath}
Indeed, it is rather easy to show (we do not provide an explicit proof here for the sake of brevity) that the mapping $\tau$ preserves the preorder $\prec$ and the equivalence relation $\approx$ (in the sense that $\alpha(x) \prec \beta(x)$ iff $\tau(\alpha(x))  \prec \tau(\beta(x))$, and $\alpha(x) \approx \beta(x)$ iff $\tau(\alpha(x))  \approx \tau(\beta(x))$). Moreover, the wff $\alpha(x) \in \phi_{TQ}(x)$ is Q--true iff the af $\tau(\alpha(x)) \in \phi_{AD}^{Q}$ is justified, which translates a semantic concept (\emph{Q--true}) defined on the language ${\mathcal L}_{TQ}(x)$ into  a pragmatic concept (\emph{justified}) defined on the pragmatic language ${\mathcal L}_{QD}^{P}$. Bearing in mind our comments at the end of Sec. 6, we can summarize these results by saying that QL can be interpreted as a theory of the notion of testability in QM from a semantic viewpoint, a theory of the notion of empirical justification in QM from a pragmatic viewpoint. The two interpretations can be connected, via the mapping $\tau$, in such a way that \emph{Q--true} transforms into \emph{justified}, which is intuitively satisfactory.

\section{Physical propositions and possible worlds}
The formal language ${\mathcal L}(x)$ introduced in Sec. 2 is exceedingly simple from a syntactical viewpoint, even if it is very useful in order to illustrate what physicists actually do when dealing with QL. Its syntactical simplicity has forced us, however, to set up a somewhat complicate semantics, in which, in particular, states are formally treated as possible worlds of a Kripke--like semantics. A less intuitive but logically more satisfactory approach should provide an extended syntactical apparatus, simplifying semantics. This could be done by enriching the alphabet of ${\mathcal L}(x)$ in two ways:

(i) adding a universal quantifier (with standard semantics);

(ii) adding the set of states as a new class of monadic predicates of ${\mathcal L}(x)$.

Let us comment briefly on these possible extensions of ${\mathcal L}(x)$. Firstly, let (i) only be introduced. Then, a family of individual propositions can be associated with the quantified wff $(\forall x) \alpha(x)$, and a proposition $p_{(\forall x) \alpha(x)}= \cap_{\rho} p_{\alpha(x)}^{\rho}$ can be associated with it. Hence, we get
\begin{displaymath}
p_{\alpha(x)}^{f}=p_{(\forall x) \alpha(x)} 
\end{displaymath}
which provides a satisfactory interpretation of the physical propositions introduced in Sec. 3 and of the related notion of certainly true.

Second, let us note that considering states as possible worlds is a common practice in QL,\cite{dcgg04} but it doesn't fit well with the standard logical interpretation of possible worlds. In order to avoid this problem, one could introduce (ii), as one of us has done, together with other authors, in several papers.\cite{g91,gs96a} In this case, states are not considered possible worlds, propositions as defined in the present paper are not propositions in the standard logical sense (rather, an `individual proposition' associated with a wff $\alpha(x)$ is the set of all states which make a sentence of the form $S(x) \rightarrow \alpha(x)$ true in a given interpretation of $x$, while a `physical proposition' is a set of `certainly yes' states which make a sentence of the form $(\forall x) (S(x) \rightarrow \alpha(x))$ true). We do not insist here on this more general scheme, and limit ourselves to observe that it is compatible with a standard Kripkean semantics, which can be enriched by introducing \emph{physical laboratories} in order to characterize the truth mode of empirical physical laws in more details and connect the notions of probability and frequency.\cite{g91,gs96a} Yet, of course, an approach of this kind would make much less direct and straightforward the interpretation of QL that we have discussed in this paper.


\end{document}